\documentstyle[12pt]{article}
\begin{document}
\setcounter{page}{0}
\topmargin 0pt
\oddsidemargin 5mm
\renewcommand{\thefootnote}{\fnsymbol{footnote}}
\newpage
\setcounter{page}{0}
\begin{titlepage}
\begin{flushright}
November 1993  \\
cond-mat/nnnmmyyy
\end{flushright}
\vspace{1.0cm}
\begin{center}
{\large {\bf NEW CRITICALITY OF 1D FERMIONS }}

\vspace{1.7cm}
{\large Michael L\"{a}ssig}

\vspace{1.1cm}
{\em Institut f\"ur Festk\"orperforschung \\
     Forschungszentrum, 52425 J\"ulich, Germany \\
     email: lassig@iff011.dnet.kfa-juelich.de}

\end{center}
\vspace{1.3cm}

\setcounter{footnote}{0}

\begin{abstract}

One-dimensional massive quantum particles (or $1+1$-dimensional random walks)
with short-ranged multi-particle interactions are studied by {\em exact}
renormalization group methods. With repulsive pair forces, such particles are
known to scale as free fermions.  With finite $m$-body forces ($m = 3,4,
\dots$), a critical instability is found, indicating the transition to a
fermionic bound state. These unbinding transitions represent new universality
classes of interacting fermions relevant to polymer and membrane systems.
Implications for massless  fermions, e.g. in the Hubbard model, are also noted.

\end{abstract}
\begin{flushright}
{\em to appear in Phys. Rev. Lett.}
\end{flushright}
\end{titlepage}

\newpage
\renewcommand{\thefootnote}{\arabic{footnote}}

Interacting quantum particles moving in one spatial dimension and imaginary
time offer a unifying description of most 2D fluctuating systems. The
trajectories of these particles represent (streched) polymers, domain walls or
interfaces, steps on surfaces, magnetic flux lines, etc. Two ensembles have to
be distinguished: (a) Vicinal surfaces \cite{VicinalSurfaces} or systems at a
2D bulk critical point (e.g. Curie point, commensurate-to-incommensurate
transition \cite{denNijs.DG}, surface reconstruction
transition~\cite{VillainVilfan.surfrec}) contain a finite density of such lines
and are described by a {\em massless} quantum field theory which is generically
isotropic and conformally invariant. (b)~Systems with only a finite number of
directed lines are ensembles of {\em massive} particles. Such systems may
exhibit critical behavior at {\em delocalization transitions} between a
low-temperature dense phase and a high-temperature dilute phase
\cite{ForgacsAl.DG}.

In the dense phase, the lines are bound to a bundle of transversal extension
$\xi_\bot$. Their relative fluctuations are thus constrained; correlations in
longitudinal direction decay on a scale $\xi_{\scriptscriptstyle \|}$. This
phase is a bound state of the quantum particles. In the dilute phase, the
lines fluctuate independently; the quantum particles are in a delocalized
state. As the transition temperature is approached from below, the length
scales $\xi_{\scriptscriptstyle \|}$ and
$\xi_{\bot} = \xi_{\scriptscriptstyle \|}^\zeta $ diverge.
These transitions are generically anisotropic; the roughness exponent $\zeta$
equals $1/2$ for temperature-driven transitions. Examples are wetting
phenomena, polymer desorption, the helix-coil transition in DNA, and unbinding
transitions of biomembrane bundles \cite{LipLeibler.memb}, which have gained
considerable experimental interest recently \cite{MutzHelfrich.memb}. Ensembles
of interacting directed lines are also important as the replica formulation of
polmers in random media \cite{Kardar.kpz}; those in turn are intimately related
to theories of nonequilibrium directed growth.

This letter aims at a systematic understanding of delocalization phenomena as
renormalized continuum field theories. An exact renormalization group (RG)
based on the short-distance algebra of the interaction vertices
\cite{ren,ALTENBERG}
reveals the existence  of a discrete series of universality classes that
represent delocalization transitions of a finite number of interacting random
walks. The possible existence of analogous massless (conformally invariant)
field theories is discussed at the end.

One-dimensional quantum particles that interact only via two-body contact
forces
define the nonlinear Schr\"odinger model, which is exactly solvable by
Bethe ansatz methods (see \cite{Thacker} for a review).  This has been applied
to unbinding transitions in refs. \cite{BurkhardtSchlottmann.unb,BETHE}.
In real systems, the interactions   are certainly more complicated than the
simple pair force of the Schr\"odinger model. Typically, the force between two
lines is screened or enhanced by the presence of further lines. Casimir-type
many-body forces (which may be screened at some scale) arise from the coupling
of the lines to the surrounding medium, e.g. a correlated fluid
\cite{Kardar.Casimir}. There is also experimental evidence for attractive
forces between steps on vicinal surfaces. When such interactions are taken into
account, the exact solvability is lost, and we are led to study their effect on
the delocalization transition by the RG.

Short-ranged  multi-particle interactions are easily shown to be irrelevant in
the sense of the RG (except 3-body forces for ``bosons'', see below). Hence
{\em weak} forces do not alter the asymptotic behavior at large distances, but
contribute only corrections to scaling.  The new universality classes describe
delocalization  transitions at {\em finite} interaction strength. In a generic
field theory, irrelevant vertices are unrenormalizable, i.e. new counterterms
are necessary at every order in perturbation theory. Remarkably enough, this
proliferation of counterterms does not take place here: the perturbation series
remains renormalizable in an $\varepsilon$-expansion although the interaction
is  irrelevant. This expansion involves analytic continuation in the number $d$
of  transversal dimensions, see eq. (\ref{em}) below.

In many of the applications above, the lines are effectively impenetrable
objects and hence do not intersect. In one dimension, this constraint on their
fluctuations is equivalent to the Pauli principle; the particles are fermions.
Particles whose trajectories are free to intersect are bosons. Repulsive
contact forces suppress intersections and hence generate a crossover from Bose
statistics to a low-energy effective Fermi statistics \cite{Thacker}.
The RG of this letter offers a unifying view on the
interplay between dynamics and statistics: delocalization transitions of bosons
and fermions fall into the same universality classes, the statistics merely
corresponds to parametrizations of the space of interactions about two
different fixed points. For the particular case of two- and three-particle
interactions, the results  are summarized in the RG flow
diagram of fig. 1 and the resulting phase diagram of fig. 2.  Depending on
these interactions, the phase transition can be governed by two distinct fixed
points, the free Bose and the ``necklace'' fixed point, which are discussed in
detail below.

We stress that all these fixed points describe ensembles of an arbitrary number
of lines; hence the critical exponents do not depend on their number. This
result agrees with ref. \cite{BurkhardtSchlottmann.unb} for the Bose fixed
point and is presumably also consistent with the extensive numerical work of
ref.  \cite{NetzLip} if the data are correctly interpreted
\cite{NumericalExponents}.

Consider a $d$-dimensional system of $p$ massive bosons coupled via forces that
decay on some microscopic scale $a$. In the continuum limit $a \to 0$, the
Hamiltonian reads
\begin{equation}
H_B^{(p)} =
\frac{1}{2} \sum_{\alpha = 1}^p  \frac{\partial^2}{\partial {\bf r}_\alpha^2}
+ \sum_{m=2}^{p} g_m \Phi_m^{(p)} \; ,
\label{Hp}
\end{equation}
where
$\Phi_2^{(p)} =
\sum_{\alpha < \beta}^p \delta^d ({\bf r}_\alpha - {\bf r}_\beta)$,
$\Phi_3^{(p)} =
\sum_{\alpha < \beta < \gamma}^p \delta^d ({\bf r}_\alpha - {\bf r}_\beta)
                               \delta^d ({\bf r}_\beta - {\bf r}_\gamma)$, etc.
are $m$-particle contact potentials.  It describes the universal behavior in
the scaling region $\xi_\bot \gg a$. In a system with long-ranged forces,
(\ref{Hp}) is still the correct continuum limit if these forces are irrelevant
in the RG, i.e. decay with a power of the distance
larger than 2. It is convenient to use a description in second quantization,
\begin{equation}
H_B = \int (\partial_{\bf r} \phi^{\scriptscriptstyle \dag} ( {\bf r}, t))
         (\partial_{\bf r} \phi ({\bf r}, t)) {\rm d}^d {\bf r} +
    \sum_{m \geq 2} g_m \Phi_m (t)  \hspace{1mm},
\label{Hbose}
\end{equation}
which is valid for an arbitrary number of lines. The operators $\phi$ and
$\phi^{\scriptscriptstyle \dag}$ obey canonical commutation relations and
\begin{equation}
\Phi_m (t) = \frac{1}{m!}
\int ( \phi^{\scriptscriptstyle \dag} ({\bf r},t) )^m
     ( \phi ({\bf r},t) )^m {\rm d}^d {\bf r}
\end{equation}
are normal-ordered $m$-particle vertices. With time as the basic scale, these
vertices have canonical dimensions
$ x_m =  (m-1) d/2 $. Hence the conjugate coupling constants $g_m$ have
dimensions
\begin{equation}
\varepsilon_m = 1 - x_m  \;.
\label{em}
\end{equation}
The vertices form the short-distance algebra \cite{OmittedTerms}
\begin{equation}
\Phi_k (t) \Phi_l (0) =
\! \sum_{ m = {\rm max} (k,l) }^{  k+l-1 } \! C_{kl}^{m} \,
| t |^{ -(k + l - m - 1) d/2 } \, \Phi_m (0) + \dots,
\label{opa}
\end{equation}
\begin{equation}
C_{kl}^{m} = \frac{ m! }{ (m - k)! (m - l)! (k + l - m)! }
             \left ( \frac{k + l - m}{2} \right )^{\! -d/2}  .
\label{Cklm}
\end{equation}

However, in order to define correlation functions of these vertices, an
infrared regularization is necessary. Here the range of each coordinate $r_i$
is compactified to a circle of radius $L^\zeta$; this regularization preserves
translational invariance in space and time.  The scale $L$ also serves to
define the dimensionless bare couplings $u_m = g_m L^{\varepsilon_m}$ and the
dimensionless free energy $ F(u_2, u_3, \dots) = L E_0(g_2, g_3, \dots;L) $ in
terms of the ground state energy $E_0$. The renormalization
consists in absorbing the singularities in the
perturbation expansion for  $F(u_2, u_3, \dots)$ into renormalized couplings
$U_m$. These singularities are encoded in the operator algebra \cite{ren}. By
virtue of (\ref{opa}), the beta function  $ \beta_m (U_2, U_3, \dots) \equiv L
\partial_L U_m $ depends only on the $U_k$ with $k \leq m$.

Hence consider first the series
$ F(u_2) = F(0) + \sum_{N=1}^{\infty} F_N u_2^N $,
where
\begin{equation}
F_N = L^{1 - N \varepsilon_2} \frac{(-1)^N}{N!} \int
      \langle \Phi_2 (t_1) \Phi_2 (t_2) \dots \Phi_2 (t_N) \rangle_L
      {\rm d} t_2 \dots {\rm d} t_N
\label{FN}
\end{equation}
and  $\langle \dots \rangle_L$ denotes connected expectation values  in the
unperturbed ground state of an arbitrary particle number sector
\cite{DavidAl.ren}. (The subsequent manipulations do not depend on the in- and
out-states but only on the short-distance structure of the correlation
functions.) In the series (\ref{FN}), a single
primitive divergence
\begin{equation}
F_2 = L^{x_2} \langle \Phi_2 \rangle_L
      C_{22}^2 L^{-\varepsilon_2}
      \int_0^L t^{-1 + \varepsilon_2} {\rm d} t + O(\varepsilon_2^0)
    = L^{x_2} \langle \Phi_2 \rangle_L \frac{1}{\varepsilon_2}
      + O(\varepsilon_2^0)
\label{sing}
\end{equation}
occurs at $\varepsilon_2 = 0$ (i.e. $ d=2$).
Hence the beta function in minimal subtraction is~\cite{beta}
\begin{equation}
\beta_2 (U_2) = \varepsilon_2 U_2 - U_2^2 \hspace{1mm}.
\label{beta2}
\end{equation}
This exact renormalizability is intimately related to the summability of the
perturbation expansion in the nonlinear Schr\"odinger model \cite{Thacker}.
Generically, (\ref{beta2}) would make sense for $\varepsilon_2 >0$, where
$U_2$ is relevant at the  Gaussian fixed point and generates a crossover to the
infrared-stable fixed point $U_2^\star = \varepsilon_2$. As an exact one-loop
equation, however, it continues  to be valid for $0 > \varepsilon_2 > -1$,
where the ultraviolet divergences in $F(u_2)$  can be absorbed in a single
counterterm.  $U_2^\star$ is then ultraviolet-stable.  In the perturbation
series for the correlation functions, singularities analogous to (\ref{sing})
at first order in $u_2$ lead to the beta functions
\begin{equation}
\beta_m (U_2, \dots, U_m) = \varepsilon_m (U_2) U_m + O(U_k U_m)
\label{betam}
\end{equation}
with $ (3 \leq k \leq m) $ and
\begin{equation}
\varepsilon_m (U_2) = \varepsilon_m - 2 C_{m2}^m U_2 + O(U_2^2)   \hspace{1mm}.
\label{ym}
\end{equation}
For $m \geq 3$, (\ref{ym}) does not terminate at first order.
In $d=1$,
however, the combined contribution from higher orders turns out to vanish
at the fixed point $U_2^\star = 1/2$, so that the infrared dimensions
resulting from
(\ref{ym}) and (\ref{Cklm}),
\begin{equation}
\bar x_m = 1 - \varepsilon_m (U_2^\star) = \frac{ m^2 - 1 }{ 2 }  \hspace{1mm},
\label{xfermi}
\end{equation}
are the exact scaling dimensions of the fermionic operators (\ref{Phibar})
below. The full beta function for $U_3$ now follows in a similar way from the
singularities in the series $F( u_2, u_3 )$ at $\varepsilon_3 = 0$.  Again, the
only  primitive singularity occurs at order $u_3^2$, and hence  (with $ C_3
\equiv C_{33}^3 $)
\begin{equation}
\beta_3 (U_2, U_3) = ( \varepsilon_3 - 3 U_2 )  U_3 - C_3 U_3^2 \hspace{1mm}.
\label{beta3}
\end{equation}

The RG flow of fig. 1 is given by (\ref{beta2}) and (\ref{beta3}) for
$d=1$. In the sequel, we discuss its three fixed points and the implications
for the phase diagram.

{\bf Free Bosons} ($U_2 = U_3 = 0$). The scale-invariant theory is
characterized by algebraic finite-size effects
$ \langle \Phi_m \rangle_L \sim L^{ - (m-1) / 2 } $.
For $g_2 < 0$, there is the well-known bound state with longitudinal
correlation length
\begin{equation}
\xi_{\scriptscriptstyle \|} (g_2) \sim |g_2|^{-2}
\label{xi2}
\end{equation}
and
$\langle \Phi_m \rangle_\infty \sim |g_2|^{m-1}$,
while $g_2 > 0$ generates the crossover to free fermions with (\ref{xi2}) now
describing the scaling of the crossover length. A repulsive three-particle
coupling $g_3 > 0$ is marginally irrelevant. For $g_2 = 0$, it leaves the
particles infrared-free, but modifies the theory
(e.g. the amplitudes $ \langle \Phi_m \rangle_L $)
on scales smaller than
\begin{equation}
\xi_{\scriptscriptstyle \|} (g_3) \sim \exp ( -1 / g_3 )     \hspace{1mm};
\label{xi3}
\end{equation}
for $g_2 ${\footnotesize $\nearrow$}$ 0$, it contributes logarithmic
corrections to scaling \cite{NumericalExponents},
e.g. $ \langle \Phi_m \rangle_\infty $\linebreak
$\sim ( g_2 / g_3 \log |g_2| )^2 $.
The marginally relevant $g_3 < 0$ leads to a bound state with (\ref{xi3}) and
$\langle \Phi_m \rangle_\infty \sim \exp ( (m-1) / 2g_3 )$; the unbinding now
takes place on the critical line $ g_2 = g_2^c (g_3) $ and is governed by the
``necklace'' fixed point described below.

{\bf Free Fermions} ($ U_2 = \varepsilon_2, U_3 = 0 $).
This fixed point describes the
limit $ g_2 \rightarrow \infty, \linebreak g_3 = 0 $,
where the particles obey the
Pauli exclusion principle. Hence the operators $\Phi_m$ vanish identically, as
follows  from the asymptotic crossover scaling of their correlation functions
given by (\ref{beta2}) and (\ref{betam}), e.g.
$ \langle \Phi_m \rangle_L (g_2) \sim
g_2^{ - m (m-1) } L^{ - (m^2 -1) / 2 } $.
Short-ranged interactions are instead described by the fermionic operators
\begin{equation}
\bar \Phi_m (t) \equiv
\frac{1}{m!} \int \prod_{i=1}^{m}
\psi^{\dag} (r + a_i) \psi (r + a_i) {\rm d} r      \hspace{1mm},
\label{Phibar}
\end{equation}
where $a_i$ are fixed microscopic distances characterizing their range. These
operators have scaling dimensions $ \bar x_m = (m^2 - 1)/2 $ \cite{xfermi} as
given by (\ref{xfermi}) and form an operator algebra of the form
(\ref{opa}). Hence for $d=1$, the  bosonic Hamiltonian (\ref{Hbose}) can be
written in the equivalent fermionic form
\begin{equation}
H_F = \int (\partial_r \psi^{\scriptscriptstyle \dag} ( r, t))
         (\partial_r \psi (r, t)) {\rm d} r +
    \sum_{m \geq 2} g_m \bar \Phi_m (t)  \hspace{1mm}.
\label{Hfermi}
\end{equation}
The fermionic RG equations are precisely of the form
(\ref{beta2}), (\ref{beta3}) with coefficients
$ \bar \varepsilon_m = 1 - \bar x_m $
and $\bar C_3$. Since all interactions (\ref{Phibar}) are irrelevant, both the
bosonic fixed
point ($ \bar U_2 = \bar \varepsilon_2 , \bar U_3 = 0 $ ) and the necklace
fixed
point ($ \bar U_2 = 0 , \bar U_3 = \bar \varepsilon_3 / \bar C_3 $ )
are ultraviolet fixed points.

{\bf Necklace Theory}
($ U_2 = \varepsilon_2, U_3 = \bar \varepsilon_3 / C_3 $).
This theory describes the critical transition between the high-temperature
phase of free fermions and the ``necklace'' bound state \cite{Fisher.walks}
that forms for $\bar g_3 < \bar g_3^c < 0$ and is named after the typical
configurations of trajectories shown in fig 2. The transition temperature
depends on the parameters $a_i$ and is nonuniversal. At this fixed point, the
three-particle coupling is relevant.  The one-loop RG predicts the
exponent
$ \varepsilon_3^{\scriptscriptstyle \triangle} =
- \bar \varepsilon_3 $ as long as
$ \bar \varepsilon_3 > -1 $. Since
$ \varepsilon_3^{\scriptscriptstyle \triangle} $
cannot become $ > \! 1 $ (this would mean  an unphysical divergence of
$ \langle \bar \Phi_3 \rangle_\infty (\bar g_3) \sim
\xi_{ \scriptscriptstyle \| }^{
     1 - \varepsilon_3}$\raisebox{1.4ex}{$\scriptscriptstyle \!\! \triangle$}
at the transition) we conclude
$ \varepsilon_3^{\scriptscriptstyle \triangle}= 1 $ for
$ \bar \varepsilon_3 < -1 $.
This is confirmed by a mapping of the necklace theory onto a particular point
of the critical line of wetting transitions \cite{FisherGelfand,NetzLip}. Hence
\begin{equation}
\xi{\scriptscriptstyle \|} (\bar g_3) \sim | \bar g_3 - \bar g_3^c |^{-1}
\hspace{1mm},
\end{equation}
and $ \langle \bar \Phi_3 \rangle_\infty (\bar g_3) $ approaches a nonuniversal
finite value as $ \bar g_3 ${\footnotesize $\nearrow$}$ \bar g_3^c $. This
implies an unusual energy balance for the necklace bound state: its kinetic
energy $E_{\rm kin}$ and potential energy $E_{\rm pot}$ remain separately
finite as the total bound state energy
$ E_{\rm kin} + E_{\rm pot} = - 1 / \xi_{\scriptscriptstyle \|} $
approaches 0, while at the bosonic transition
$ E_{\rm kin} \simeq - E_{\rm pot} / 2  \simeq 1 / \xi_{\scriptscriptstyle
\|}$.

{}From the foregoing RG analysis it transpires that the fixed
point $\bar U_3^\star$ is just the first member of a whole family of fermionic
necklace theories represented by fixed points $\bar U_m^\star$ of the higher
multi-particle interactions $\bar \Phi_m$. Thus the interplay between
attractive and repulsive forces generates a rich scenario of universality
classes of interacting directed walks. A detailed understanding of their
correlation functions and the various crossover phenomena is within the reach
of these RG methods but beyond the scope of this letter. The Bethe ansatz
yields the correct asymptotic scaling if and only if the Hamiltonian
(\ref{Hbose}) is in the universality class of the free Bose or Fermi fixed
point. Whether analogous methods of exact solution exist for the higher
fixed points $\bar \Phi_m$ is an open question.

A further interesting question is the existence of analogous fixed points for
theories of massless relativistic fermions, where isotropy is restored through
particle-antiparticle processes. The simplest case is the critical point of the
2D Ising model, a theory  of free Majorana fermions with action
$ S = \int ( \psi_+ \partial_- \psi_+ +
      \psi_- \partial_+ \psi_- ) {\rm d}^2 {\bf r} $
in terms of the chiral components $\psi_+$ and $\psi_-$. The
lowest-dimensional scalar interaction
that is local in the Fermi fields is the irrelevant normal-ordered
$4$-particle vertex
$ : \!\! \psi_+ \psi_+ \psi_+ \psi_+ \psi_- \psi_- \psi_- \psi_- \!\! : \;
= T_+ T_- $ (where $T_+$ and $T_-$ denote the components of the stress
tensor). This interaction is known to be integrable and to generate a
crossover whose ultraviolet ``necklace'' fixed point is the tricritical Ising
model \cite{KMS.Tim/AlZamolodchikov.Tim}. Thus it is tempting to associate the
hierarchy of multi-particle interactions with the famous series of minimal
conformal field theories \cite{BPZ}.
Many other applications involve Dirac fermions with (marginal) local pair
interactions. Examples are the ubiquitous Gaussian model with the fermionic
action
$ S = \int ( \bar \psi_+ \partial_- \psi_+ +
             \bar \psi_- \partial_+ \psi_- +
      g_2 \bar \psi_+ \psi_+ \bar \psi_- \psi_- ) {\rm d}^2 {\bf r}  $,
or the Hubbard model, a theory of two Dirac fermions coupled by similar pair
forces (that is relevant to roughening of reconstructed surfaces
\cite{BalentsKardar.rec}). In these cases,
the effects of the higher interactions (e.g. $ T_+ T_- $) are unknown, but
since they have a self-coupling in the operator algebra, they
are likely to generate similar transitions to massive strong-coupling phases.
These multicritical Dirac theories would correspond to conformal field theories
with central charge $c > 1$.

I am grateful to T.W. Burkhardt, H. Kinzelbach, R. Lipowsky, and R. Netz for
useful discussions and comments.

%BIBLIO

% \newpage

\vspace{1cm}

{\bf Figure Captions}

\begin{enumerate}

\item
RG flow diagram. $U_2, U_3$ and $\bar U_2, \bar U_3$ denote renormalized
two- and three-particle couplings about the fixed points of  free bosons ($
\circ $) and free fermions  ($ \bullet $), respectively.  The
transition is governed by the Bose fixed point for $U_3 \geq 0$ and by the
necklace fixed point ($ \scriptsize \triangle $)  for $ U_3 < 0 $.

\item
Phase diagram in the bare couplings $u_2,
u_3 $ of the bosonic theory. Typical world-line configurations:
free bosons ($ u_2 = u_3 = 0 $),
free fermions ($ u_2 \rightarrow \infty,$\linebreak$ u_3 = 0 $),
bound state for $ u_2 < 0 $ and $ u_3 > 0 $,
bound state for $ u_2 > 0 $ and $ u_3 < 0 $.

\end{enumerate}


\begin{thebibliography}{99}
\bibitem{NumericalExponents}
The authors of ref. \cite{NetzLip} report a
violation of (\ref{xi2}), but their numerical data can probably be accounted
for by corrections due to a repulsive $3$-particle coupling  implicitly
present in their system.
\bibitem{OmittedTerms}
The $\Phi_m$ are normalized such that $C_{22}^2 = 1$.
The terms omited on the r.h.s. contain gradients in space or time or are
regular as $ |t-t'| \rightarrow 0$.
\bibitem{xfermi}
This was first shown in ref. \cite{Fisher.walks} without the explicit use of
free fermions. The easiest derivation is to consider the expectation value in a
$m$-particle state
$ \langle \chi | \bar \Phi_m | \chi \rangle_L =
\int | \chi ( r + a_1, \dots, $\linebreak$ r + a_m ) |^2 {\rm d} r $,
which scales as
$ L^{ - (m-1) / 2 } | \prod_{i<j} \epsilon_{ij} |^2 \sim L^{- \bar x_m} $
with $ \epsilon_{ij} \equiv (a_i - a_j) L^{-1/2} $,
due to the antisymmetry of the fermionic wave function.
\end{thebibliography}
\end{document}